\documentclass[aps,10pt,prb,twocolumn,groupedaddress,amsmath,amssymb,longbibliography,floatfix]{revtex4-2}
\usepackage{graphicx}% Include figure files
\usepackage{dcolumn}% Align table columns on decimal point
\usepackage{bm}% bold math
\usepackage{physics}
\usepackage{hyperref}
\usepackage{xcolor}
\usepackage{orcidlink}
\usepackage[T1]{fontenc}
\usepackage{todonotes}
\DeclareGraphicsExtensions{.eps,.pdf,.png,.jpg}

\definecolor{Blue}{RGB}{31, 119, 180}

\definecolor{Orange}{RGB}{255, 127, 14}

\definecolor{Green}{RGB}{44, 160, 44}

\definecolor{MeasRed}{RGB}{214, 39, 40}

\definecolor{TpaPurple}{RGB}{148, 103, 189}

\newcommand{\Us}{\ensuremath{U_{\text{s}}}}
\newcommand{\Up}{\ensuremath{U_{\text{p}}}}
\newcommand{\Vs}{\ensuremath{V_{\text{s}}}}
\newcommand{\Vp}{\ensuremath{V_{\text{p}}}}
\newcommand{\Phis}{\ensuremath{\Phi_{\text{s}}}}
\newcommand{\Phip}{\ensuremath{\Phi_{\text{p}}}}

\newcommand{\rs}{\ensuremath{r_{\text{s}}}}
\newcommand{\rp}{\ensuremath{r_{\text{p}}}}
\newcommand{\ts}{\ensuremath{t_{\text{s}}}}
\newcommand{\tp}{\ensuremath{t_{\text{p}}}}

\newcommand{\epsr}{\ensuremath{\epsilon_{\text{r}}}}
\newcommand{\mur}{\ensuremath{\mu_{\text{r}}}}

\newcommand{\sigmasurf}{\ensuremath{\sigma_\text{s}}}
\newcommand{\sigmabar}{\ensuremath{\bar{\sigma}}}
\newcommand{\dsigma}{\ensuremath{\Delta \sigma}}
\newcommand{\dsigmasat}{\ensuremath{\Delta \sigma_{\text{sat}}}}
\newcommand{\dsigmatpa}{\ensuremath{\Delta \sigma_{\text{TPA}}}}
\newcommand{\dsigmasurf}{ \Delta \sigma_{\text{s}}}

\newcommand{\E}{\ensuremath{\mathcal{E}}}
\newcommand{\Esat}{\ensuremath{\mathcal{E}_{\text{sat}}}}

\newcommand{\chif}{\chi_{F}}
\newcommand{\chiE}{\ensuremath{\chi_\mathcal{E}}}

\newcommand{\fsat}{F_{\text{sat}}}
\newcommand{\ftpa}{\ensuremath{F_{\text{TPA}}}}
\newcommand{\funit}{mJ/cm\textsuperscript{2}}

\newcommand{\kcsixty}{\ensuremath{\text{K}_{3}\text{C}_{60}}}

\newcommand*\squeezespaces[1]{% %% <- #1 is a number between 0 and 1
  \thickmuskip=\scalemuskip{\thickmuskip}{#1}%
  \medmuskip=\scalemuskip{\medmuskip}{#1}%
  \thinmuskip=\scalemuskip{\thinmuskip}{#1}%
  \nulldelimiterspace=#1\nulldelimiterspace
  \scriptspace=#1\scriptspace
}
\newcommand*\scalemuskip[2]{%
  \muexpr #1*\numexpr\dimexpr#2pt\relax\relax/65536\relax
} %% <- based on  https://tex.stackexchange.com/a/198966/156366

\begin{document}

\title{Nonlinear photoconductivity in pump-probe spectroscopy. I. Optical coefficients}
\author{Leya Lopez \orcidlink{0000-0003-2452-803X}}
\author{Derek G.~Sahota \orcidlink{0000-0003-1730-0766}}
\author{J.~Steven Dodge\orcidlink{0000-0003-3219-7314}}
\email{jsdodge@sfu.ca}
\affiliation{Department of Physics, Simon Fraser University, V5A~1S6, Canada}

\date{\today}

\begin{abstract}
We analyze the optical pump-probe reflection and transmission coefficients when the photoinduced response depends nonlinearly on the incident pump intensity.
Under these conditions, we expect the photoconductivity depth profile to change shape as a function of the incident fluence, unlike the case when the photoinduced response is linear in the incident intensity.
We consider common optical nonlinearities, including photoconductivity saturation and two-photon absorption, and we derive analytic expressions for the photoconductivity depth profile when one or more is present.
We review the theory of the electromagnetic transmission and reflection coefficients in a stratified medium, and we derive general expressions for these coefficients for a medium with an arbitrary photoconductivity depth profile.
For several photoconductivity profiles of importance in pump-probe spectroscopy, we show that the wave equation can be transformed into one of three standard differential equations---the Bessel equation, the hypergeometric equation, and the Heun equation---with analytic solutions in terms of their associated special functions.
From these solutions, we derive exact analytic expressions for the optical coefficients in terms of the photoconductivity at the optical interface, and we discuss their limiting forms in various physical limits.
Our results provide a systematic guide for analyzing pump-probe measurements over a wide range of pump intensities, and establishes a framework for constraining the systematic uncertainty associated with nonlinear photoconductivity profile distortion.
\end{abstract}

\maketitle

\section{Introduction}
\label{sec:Introduction}
Optical pump-probe spectroscopy uses ultrafast light pulses to investigate how solids evolve in response to optical excitation, enabling the study of dynamics associated with electronic, magnetic, and lattice degrees of freedom at their fundamental time scales~\cite{jepsen2011, ulbricht2011, basov2011}.
The advent of powerful laser sources capable of delivering intense optical pulses has expanded the opportunities for pump-probe spectroscopy to investigate nonequilibrium states of matter~\cite{ basov2017, delatorre2021}.
However, in these experiments, the measured optical coefficients, such as the reflection and transmission amplitudes, represent an averaged response over the entire photoexcited region, whereas the response functions of interest, such as the conductivity or permittivity, are local properties that can vary both temporally and spatially within the region.
Consequently, we can not use pump-probe measurements to determine local pump-induced conductivity changes without a theoretical model for the photoexcitation depth profile.

A common approach to this problem is to assume that the pump excitation produces a local response that decays exponentially with depth, as it would in the linear response regime.
While we can justify this assumption when the incident intensity is sufficiently low, it fails as the intensity increases because optical nonlinearity becomes important.
(Since photoconductivity is itself a nonlinear optical phenomenon, unless otherwise indicated we use the terms \emph{nonlinear} and \emph{nonlinearity} here to refer to the relationship between the local photoconductivity at any point in the medium and the incident pump power or fluence.)
Ignoring this nonlinearity can lead to systematic errors in the inferred photoconductivity that can be as large as the response itself~\cite{dodge2023}.
To address this problem, we have developed an analysis framework that can account for both low-intensity excitation conditions and the nonlinear effects that emerge at higher pump intensity.
It is based on a family of photoexcitation profiles that depend on the incident fluence, with two additional parameters that define characteristic scales for saturable and two-photon nonlinearities.
A notable feature of this family of profiles is that it supports analytic solutions to the wave equation in terms of known special functions.
We derive these solutions and analyze them in detail here.
In subsequent work, we will describe how to use this family of profiles to improve the systematic uncertainty in pump-probe measurements in the high-intensity limit.

To motivate our discussion, we note that our interest in this topic emerged from earlier efforts to determine the photoinduced conductivity from pump-probe measurements at high pump fluence~\cite{petersen2017, sahota2019, dodge2023}.
Initially, we found that the pump-probe response in insulating cuprates shows evidence of saturation as the pump fluence reaches about 1~\funit~\cite{petersen2017, sahota2019}.
This observation prompted us to consider the influence of this nonlinearity on the photoexcitation profile~\cite{sahota2019}, which subsequently led us to consider whether a similar nonlinearity could affect the optical evidence for photoinduced superconductivity~\cite{cavalleri2018}.
By reanalyzing the evidence for photoinduced superconductivity in $\kcsixty$, we found strong evidence that it is distorted by a saturable nonlinearity~\cite{dodge2023}.
We argued that most of the optical evidence for photoinduced superconductivity suffers from the same systematic error.
Although our analysis focused on the example of photoinduced superconductivity, we expect similar problems to emerge in any pump probe experiment at sufficiently high pump intensity.
This recognition led us to the present work, where we aim to lay the foundation for an analysis protocol that can account for nonlinearities more generally.

We discuss how the photoconductivity depends on depth in the presence of common optical nonlinearities in Sec.~\ref{sec:photoconductivity-models}.
This section elaborates on our previously published work on saturation and two-photon absorption (TPA) nonlinearities~\cite{dodge2023}, and it develops an expression that can account for both nonlinearities simultaneously.
We discuss limiting forms of this expression and relate its general features to profiles expected for other types of nonlinearity.
In Sec.~\ref{sec:general-formalism}, we review the general problem of solving the Maxwell equations in stratified media.
We discuss common solution methods, and we derive expressions for the optical coefficients that are valid for an arbitrary photoconductivity depth profile.
This prepares us for Sec.~\ref{sec:analytic-solutions}, where we derive exact solutions to the wave equation for the photoconductivity depth profiles introduced in Sec.~\ref{sec:photoconductivity-models}.
We also derive exact analytic expressions for the reflection and transmission amplitudes under a variety of conditions that are relevant for pump-probe spectroscopy.
We summarize and conclude in Sec.~\ref{sec:conclusion}.

\section{Models for photoconductivity depth profile}
\label{sec:photoconductivity-models}
Photoexcitation changes the local conductivity of the medium by depositing energy, which is then redistributed across various degrees of freedom within the material.
At low pump intensity, we expect the absorbed energy density at every location in the sample to be proportional to the incident pump fluence, and the change in local conductivity to scale with the absorbed energy density.
However, as the pump intensity increases, nonlinear effects are likely to emerge in the pump absorption process, the medium response, or both.
In this work, we focus on two common nonlinearities: one  where the pump absorption is nonlinear due to TPA or excited-state absorption (ESA), and another where the local conductivity becomes nonlinear as it saturates with the absorbed energy density.
We consider these two cases individually and we also consider a third case where we combine these two nonlinearities.

We start with two important simplifying assumptions.
First, we treat the photoexcited medium as if it were in a quasiequilibrium steady state, where the photoinduced changes occur on a timescale longer than the pulse duration of the probe~\cite{ulbricht2011}.
This assumption allows us to focus solely on the spatial variation of the local photoconductivity and its relationship to the incident pump fluence.
Also, we assume that the characteristic relaxation time of the material is shorter than any experimental timescale, in particular the pump-probe delay.
This allows us to derive the optical reflection and transmission coefficients by treating the photoexcited medium in the quasistatic limit and assuming a linear interaction with the probe, even though the underlying photoconductivity response is fundamentally a nonlinear optical phenomenon~\cite{orenstein2015}.

By focusing on these two nonlinearities, TPA and saturation, our framework extends the more restrictive conventional analysis to include two of the most common nonlinearities, while keeping the analysis procedure tractable.
As we show later, these nonlinearities exhibit contrasting dependencies on fluence, and by combining them, we develop a model that can describe a wider range of nonlinearities in the experimental regime.
Moreover, these models can be adapted to account for other types of nonlinearities, which we will discuss in later sections.

A key advantage of our approach is that the wave equation corresponding to these nonlinearities, as well as their combined form, have analytic solutions in terms of known special functions.
As discussed in Sec.~\ref{sec:general-formalism}, the standard approach to determine the optical coefficients of a stratified medium requires the solution for the wave that propagates into the medium.
In general, this must be done numerically, which can require some care if the medium is absorbing and the forward-propagating wave decays exponentially into it.
However, when exact analytic solutions are available for the medium, we can relate the optical coefficients directly to the response functions at the surface, avoiding the need to compute the fields at any other location.

In the following subsections, we introduce the photoconductivity depth profile models that we study in this paper.
For completeness and to establish notation, we begin our discussion by introducing the linear photoconductivity depth profile in Sec.~\ref{sec:linear-photoconductivity}.
In Sec.~\ref{sec:saturable-photoconductivity}, we introduce the saturable photoconductivity depth profile, where we assume that the pump absorption remains linear with the pump fluence, but the photoconductivity saturates with the absorbed energy density.
Section~\ref{sec:tpa-photoconductivity} focuses on the TPA profile, where the pump absorption process is nonlinear due to two photon absorption while the photoconductivity is linear with the absorbed energy density.
In Sec.~\ref{sec:combined-photoconductivity}, we introduce the combined profile, where both the pump absorption and the response are nonlinear due to TPA and saturation.
Finally, in subsection Sec.~\ref{sec:relevant-photoconductivity}, we discuss other common nonlinearities and relate their behavior to the depth profiles we developed in earlier sections.
 
\subsection{Exponential depth profile}
\label{sec:linear-photoconductivity}
As noted earlier, at sufficiently low pump intensity we can assume that the local conductivity depends linearly on the local energy density, $\E$.
We express this mathematically as $\sigma(z, \E) =  \sigmabar + \chiE \E(z)$, where $z$ is the depth from the surface, $\sigmabar$ is the equilibrium conductivity,  \mbox{$\chiE = \left.d\sigma/d\E\right|_{z=0}$} is the complex susceptibility of the conductivity to the absorbed energy density at the surface, and $\sigma$, $\sigmabar,$ and $\chiE$ all have implicit frequency dependence.
If the nonlinear absorption of the pump beam is weak, we may assume further that \mbox{$\E = \E_0 e^{-\alpha z}$}, where $\alpha$ is the pump attenuation coefficient.
We may then write $\sigma$ as a function of both $z$ and the incident fluence, $F$,
\begin{equation}
    \sigma(z, F) = \sigmabar + \chif F e^{-\alpha z},
\label{eq:sigma-exp-profile-fluence}
\end{equation}
where \mbox{$\chif = \lim_{F\rightarrow 0}d\sigma/dF$}, which we call the surface photosusceptibility.
Setting $z = 0$ in Eq.~\eqref{eq:sigma-exp-profile-fluence}, we write the surface photoconductivity as
\begin{equation}
    \dsigmasurf (F) \equiv \sigma(z=0, F) - \sigmabar =  \chif F.
\label{eq:surfsigma-exp-profile}
\end{equation}
It is important to recognize that $\sigma(z, F)$ and $\dsigmasurf (F)$ are not a true linear response coefficients, in the sense that they describe a third-order optical nonlinearity that relates the current density at the surface to both the probe field amplitude and the pump fluence~\cite{orenstein2015}.
By introducing the third-order surface photosusceptibility $\chif$, we make this fluence dependence explicit.
This quantity has the added advantage that it retains a well-defined interpretation when $\sigma(z, F)$ has a nonlinear dependence on $F$, as we discuss below.

\subsection{Saturable profile}
\label{sec:saturable-photoconductivity}

As the pump fluence increases, we expect the energy distribution to change among the different degrees of freedom in the system.
To estimate of the energy deposited in the system in a typical experiment, consider a medium with a 50\% absorptance and the pump attenuation length is 100~nm that is that is photoexcited with an incident pump fluence of 1~\funit.
The absorbed energy density is then approximately 0.3~eV/nm\textsuperscript{3}, which is large enough to expect internal degrees of freedom to saturate.
For example, in the Heisenberg model of a magnetic system, the energy absorption is limited by \mbox{$\mathcal{E}_\text{max}\propto JS^2$}, where $J$ is the exchange energy and $S$ is the total spin.
In cuprates, the magnetic degrees of freedom can store a maximum of about 0.5~eV/nm\textsuperscript{3} per unit cell, which is comparable to the 0.3~eV/nm\textsuperscript{3} in our example.
While phonons are not constrained in the same way, we may expect that anharmonic coupling will allow a phonon to redistribute its energy to other degrees of freedom more rapidly as its energy increases.
If a saturating degree of freedom makes the dominant contribution to the photoconductivity, we expect the photoconductivity to saturate as the fluence increases.
This is observed in a variety of experiments~\cite{vanderhoef2014, lui2014, tomadin2018, jensen2014, sahota2019, meng2015}.

We model this behavior by first expressing the photoconductivity as a function of the energy density,
\begin{align}
    \dsigma (z, \E; \Esat) &=  \chiE \frac{\E (z, F)}{1 + \E (z, F)/\Esat},
\label{eq:sat_sigma_energy}
\end{align}
where $\Esat$ is the energy density at which $\dsigma$ saturates and \mbox{$\chiE = \lim_{\E \rightarrow 0}d\sigmasurf/d\E$}, the susceptibility of the surface photoconductivity to the absorbed energy density.
Assuming that the pump absorption remains linear with fluence, \mbox{$\E \propto F\exp(- \alpha z)$}, and the saturable conductivity depth profile is
\begin{align}
\sigma(z, F; \fsat) &= \sigmabar + \chif \fsat \frac{\left(F/\fsat\right)  e^{-\alpha z}}{1 + \left( F/\fsat \right) e^{-\alpha z}}.
\label{eq:sigma-sat-profile}
\end{align}
Here, \mbox{$\fsat = \mathcal{E}_{\text{sat}}/[\alpha(1 - R_{\text{p}})]$}, where $R_{\text{p}}$ is the pump reflection coefficient, is a parameter that sets the fluence scale of saturation.
As \mbox{$F/\fsat \rightarrow 0$}, we recover the exponential model that describes the local photoconductivity in the linear regime.
As \mbox{$F/\fsat \rightarrow \infty$}, the surface photoconductivity approaches the limiting value \mbox{$\dsigmasat = \chif \fsat$}.
Setting $z = 0$, the surface photoconductivity for a saturable depth profile has the form
\begin{equation}
    \sigmasurf (F; \fsat) = \sigmabar + \chif \fsat \frac{ F/\fsat}{1 + F/\fsat}, 
\label{eq:surfsigma-sat-profile}
\end{equation}
which exhibits a sublinear relationship with the pump fluence.

\begin{figure}[tbp]
\begin{center}
\includegraphics[width=\columnwidth]{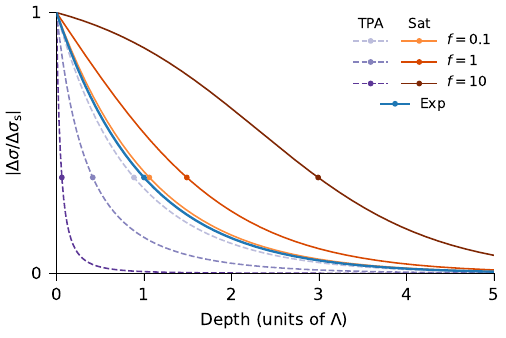}
\caption{Normalized photoconductivity of linear and nonlinear media as a function of depth, in units of the pump penetration depth $\Lambda = 1/\alpha$.
Markers on each curve indicate the $1/e$-depth.
The photoexcition profile narrows with increasing fluence $F = \ftpa f$ in a medium with a TPA nonlinearity and broadens with increasing fluence $F = \fsat f$ in a medium with a saturable nonlinearity (Sat).
The depth profile for a linear medium (Exp) is independent of fluence.}
\label{fig:sat-tpa-profile}
\end{center}
\end{figure}

The orange curves in Fig.~\eqref{fig:sat-tpa-profile} show $\dsigma(z, F; \fsat)$ given by Eq.~\eqref{eq:sigma-sat-profile} for different values of pump fluence, with each curve normalized to its value at the surface.
The blue curve shows the exponential photoconductivity depth profile given by Eq.~\eqref{eq:sigma-exp-profile-fluence}, and the violet curves show the photoconductivity depth profile for a TPA nonlinearity, which will be discussed in detail in the Sec.~\ref{sec:tpa-photoconductivity}.
While the functional form of the exponential model is independent of the fluence, the shape of the saturable profile varies strongly with it.
As the fluence increases, the local conductivity saturates more quickly at the surface than in the interior, where the absorbed energy density is lower.
This causes the profile to broaden, as the region of saturated photoconductivity extends deeper into the material.

\subsection{TPA profile}
\label{sec:tpa-photoconductivity}

Unlike the saturable profile, the TPA profile becomes shallower with increasing fluence, as shown in Fig.~\ref{fig:sat-tpa-profile}.
This happens because as the fluence increases, the likelihood of carrier excitation is higher at the surface, where the energy is absorbed faster before it can penetrate deeper into the medium.
This confines the effective region of excitation more tightly to the surface.

To derive the TPA photoconductivity depth profile, we begin by calculating the absorbed energy density in a TPA process.
During a TPA process, an electron is excited from the ground state via the simultaneous absorption of two pump photons~\cite{boyd2008}.
The spatial variation of the pump intensity, which accounts for both the linear absorption and TPA, can be modeled using the differential equation
\begin{align}
\frac{dI}{dz} = - \alpha I - \beta I^{2},
\label{eq:tpa-ode}
\end{align}
where $\beta$ is the TPA coefficient.
The resulting solution for the pump intensity is
\begin{align}
I(z, t) = \frac{I_{\text{s}} (t)e^{- \alpha z}}{1 + \beta I_{\text{s}}/\alpha \left(1 - e^{- \alpha z}\right)},
\end{align}
where \mbox{$I_{\text{s}}(t) = I(z = 0^{+},t)$}.
Assuming a constant pump illumination across the probe surface and a rectangular temporal profile with pulse width $\tau_{\text{p}}$, the absorbed energy density is
\begin{align}
    \E(z, F) &= - \int_{-\tau_{p}/2}^{\tau_{p}/2} dt \frac{dI}{dz} \nonumber \\ &=  (1 - R_{\text{p}}) \alpha \ftpa \frac{F/\ftpa (1 + F/\ftpa)e^{-\alpha z}}{\left[1 + F/\ftpa(1 - e^{-\alpha z})\right]^{2}}, 
\label{eq:energy_density_TPA}
\end{align}
where \mbox{$\ftpa = (\alpha \tau_{\text{p}}/\beta)/(1 - R_{\text{p}})$} represents the characteristic fluence scale for the onset of the TPA process.
Assuming that the photoconductivity is linear in the absorbed energy density, \mbox{$ \dsigma \propto \E (z)$}, we obtain
\begin{widetext}
\begin{equation}
\sigma(z, F; \ftpa) = \sigmabar + \chif \ftpa  \frac{\left(F/\ftpa)\right (1 + F/\ftpa) e^{-\alpha z}}{\left[1 + F/\ftpa(1-e^{-\alpha z})\right]^2}.
\label{eq:sigma-tpa-profile}
\end{equation}
\end{widetext}

Setting \mbox{$z = 0$} in Eq.~\eqref{eq:sigma-tpa-profile}, we obtain the surface photoconductivity,
\begin{equation}
\squeezespaces{0.5}
\dsigmasurf(F; \ftpa) = \chif \ftpa \left( F/\ftpa \right ) \left(1 + F/\ftpa\right).
\label{eq:surfsigma-tpa-profile}
\end{equation}
As with the saturable profile, we can define a characteristic photoconductivity scale in Eq.~\eqref{eq:surfsigma-tpa-profile}, \mbox{$\dsigmatpa = \chif \ftpa$}, although its significance is different.
When $F\ll\ftpa$, the surface photoconductivity is approximately linear in fluence, with $\dsigmasurf \approx \chif F$.
As $F$ increases, this dependence crosses over to \mbox{$\dsigmasurf \approx \chif F^2/\ftpa$}, rises to \mbox{$\dsigmasurf = 2\dsigmatpa$} at $F=\ftpa$, and diverges as $F\rightarrow\infty$.
Time-resolved THz measurements in ZnO show a qualitatively similar dependence of the carrier density on fluence for blue excitation wavelengths~\cite{baxter2009}.

\subsection{Combined saturable-TPA profile}
\label{sec:combined-photoconductivity}
In many practical situations, we expect both of the nonlinearities described in Sec.~\ref{sec:saturable-photoconductivity} and Sec.~\ref{sec:tpa-photoconductivity} to be relevant.
To describe this, we substitute $\E$ in Eq.~\eqref{eq:energy_density_TPA} into the expression for $\dsigma$ in Eq.~\eqref{eq:sat_sigma_energy}, we derive the local photoconductivity for the combined profile as
\begin{widetext}
\begin{equation}
    \sigma (z, F; \fsat, \ftpa) = \sigmabar + \chif\fsat \frac{\left(F/\fsat \right) \left(1 + F/\ftpa \right)e^{- \alpha z}}{\left[1 + (F/\ftpa)\left(1 - e^{-\alpha z}\right)\right]^{2} + (F/\fsat) \left(1 + F/\ftpa\right)e^{-\alpha z}},
\label{eq:sigma-tpa-sat-profile}
\end{equation}
\end{widetext}
which is now described by the two free parameters $\fsat$ and $\ftpa$.
Note that this profile has the same characteristic scale for the photoconductivity, $\dsigmasat = \chif\fsat$, as the saturable profile in Eq.~\eqref{eq:sigma-sat-profile}, which cuts off the divergence that occurs in Eq.~\eqref{eq:sigma-tpa-profile} as $F\rightarrow\infty$.
Setting \mbox{$z = 0$}, the surface photoconductivity associated with Eq.~\eqref{eq:sigma-tpa-sat-profile} is
\begin{widetext}
\begin{equation}
    \dsigmasurf(F; \fsat, \ftpa) = \chif \fsat  \frac{\left(F/\fsat \right) \left(1 + F/\ftpa \right)}{1 +  (F/\fsat)\left(1 + F/\ftpa \right)}.
\label{eq:surfsigma-comb-profile}
\end{equation}
\end{widetext}

\begin{figure}[tbp]
\begin{center}
\includegraphics[width=\columnwidth]{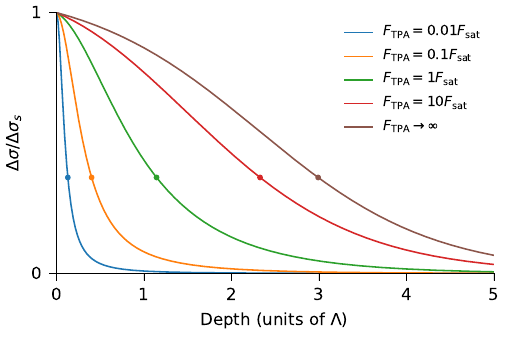}
\caption{Normalized photoconductivity as a function of depth for the combined profile at $F = 10F_{\text{sat}}$ for different values of $F_{\text{TPA}}$.
Markers on each curve indicate the $1/e$-depth.}
\label{fig:combined_depth_profile}
\end{center}
\end{figure}

Equation~\eqref{eq:surfsigma-comb-profile} describes a two-parameter family of photoconductivity profiles that can now describe both the sublinear fluence dependence of the saturable profile and superlinear dependence of the TPA profile.
Figure~\ref{fig:combined_depth_profile} shows \mbox{$\dsigmasurf(F; \fsat, \ftpa)$} for different ratios of $\ftpa/\fsat$, with each curve normalized to its value at the surface.
The depth profiles show opposing trends at the two limits of $\ftpa/\fsat$ ratios, approaching the saturable profile as \mbox{$\ftpa/\fsat \rightarrow 0$} and the TPA profile as \mbox{$\ftpa/\fsat \rightarrow \infty$}.
In the limit \mbox{$\ftpa \rightarrow 0$} (not shown), the photoconductivity drops discontinuously from $\dsigma = \dsigmasat$ at the surface to zero just inside the medium.

\subsection{Other profiles with nonlinear fluence dependence}
\label{sec:relevant-photoconductivity}
As we have described previously~\cite{dodge2023}, we can use the profiles described in Secs.~\ref{sec:saturable-photoconductivity}--\ref{sec:combined-photoconductivity} to gain insight into the behavior expected with other nonlinearities, which we review briefly here.
In Sec.~\ref{sec:saturable-scattering}, we show that Eq.~\eqref{eq:sigma-sat-profile} can be extended to the case of a conductor in which photoexcitation induces a saturable change in the carrier momentum relaxation rate.
In Sec.~\ref{sec:heating}, we discuss how heating can induce changes that are qualitatively similar to those described by Eq.~\eqref{eq:sigma-sat-profile}.
In Sec.~\ref{sec:esa}, we discuss the near-equivalence of Eq.~\eqref{eq:sigma-tpa-profile} and the profile obtained for ESA.

\subsubsection{Saturable momentum relaxation rate}
\label{sec:saturable-scattering}
To derive Eq.~\eqref{eq:sigma-sat-profile}, we assumed a saturable dependence of $\dsigma$ on $F$.
Here, we assume instead that the local carrier momentum relaxation rate of a conducting medium has an analogous dependence, as has been observed, for example, in high-resistivity silicon~\cite{meng2015}.
We describe the local photoconductivity using a Drude-Lorentz model,
\begin{equation}
\sigma(\omega, z, F; F_{\gamma}) = \frac{\epsilon_{0} \omega^{2}_\text{p}}{\gamma(z, F) - i\omega} + \sigma_{\text{b}}(\omega),
\label{eq:mob_sat_drude}
\end{equation}
where $\omega_\text{p}$ is the plasma frequency of the mobile carriers, $\sigma_{\text{b}} (\omega)$ is the conductivity associated with bound charge, and the momentum relaxation rate $\gamma$ saturates with fluence at the characteristic scale $F_{\gamma}$,
\begin{equation}
\gamma (z, F; F_{\gamma}) = \bar{\gamma} + \Delta \gamma \frac{(F/F_{\gamma})e^{-\alpha z}}{1 + (F/F_{\gamma})e^{-\alpha z}}.
\label{eq:sat_scattering}
\end{equation}
Here, $\bar{\gamma}$ is the equilibrium relaxation rate and the saturated change in this rate is \mbox{$\Delta \gamma = \gamma_{\text{sat}} - \bar{\gamma}$}, where $\gamma_{\text{sat}}$ is the saturated rate.
Substituting Eq.~\eqref{eq:sat_scattering} into Eq.~\eqref{eq:mob_sat_drude}, we can write the changes in the local photoconductivity as
\begin{equation}
\dsigma (\omega, z, F; F_{\gamma}) =  \frac{\epsilon_{0} \omega^{2}_{p}}{\bar{\gamma} - i\omega}\frac{\bar{\gamma} - \gamma_{\text{sat}}}{\gamma_{\text{sat}} - i \omega}\frac{\tilde{n}}{1 + \tilde{n}},
\label{eq:sat_scattering_profile}
\end{equation}
where 
\begin{equation}
\tilde{n} = \left(F/F_{\gamma}\right)\frac{\bar{\gamma} - \gamma_{\text{sat}}}{\gamma_{\text{sat}} - i \omega}e^{-\alpha z}.
\label{eq:sat-mobilit-tilde-n}
\end{equation}
With the substitutions
\begin{align}
\frac{\epsilon_{0} \omega^{2}_{p}}{\bar{\gamma} - i\omega}\frac{\bar{\gamma} - \gamma_{\text{sat}}}{\gamma_{\text{sat}} - i \omega} &\rightarrow \chif\fsat, & F_{\gamma}\frac{\gamma_{\text{sat}} - i \omega}{\bar{\gamma} - \gamma_{\text{sat}}} &\rightarrow \fsat,
\end{align}
we recover Eq.~\eqref{eq:sigma-sat-profile}, so despite the differences in the underlying physical models, mathematically, these profiles are equivalent.

\subsubsection{Heating-induced nonlinearity}
\label{sec:heating}
Heating is one of the most commonly considered nonlinearities in pump-probe spectroscopy.
In this case, the energy deposited via photoexcitation raises the local temperature of the medium, which will typically cause the photoconductivity to show a sublinear dependence on fluence~\cite{niwa2019, tomadin2018, jensen2014}.
This can be understood within the Debye model.
If the specific heat has the usual phonon dependence, $C_{\text{v}} \propto T^{3}$, then the temperature change associated with a given $\E$ will be $\Delta T \propto \E^{1/4}$.
Assuming $\dsigma \propto \Delta T$, this yields $\dsigma(F)\propto F^{1/4}$.
Although this does not saturate with $F$ as Eq.~\eqref{eq:sigma-sat-profile} does, the dependence is still strongly sublinear, so the behavior over any finite fluence range will be qualitatively similar.

\subsubsection{ESA and saturable absorption}
\label{sec:esa}
Finally, we consider the nonlinearities due to ESA and saturable absorption (SA).
Both nonlinearities are described by the differential equation~\cite{tutt1993}
\begin{equation}
\frac{dF}{dz} = -\alpha F - \gamma F^{2},
\label{eq:esa-ode}
\end{equation}
where $\gamma$ represents the ESA coefficient when $\gamma>0$ and the SA coefficient when $\gamma<0$.
The form of this equation is identical to Eq.~\eqref{eq:tpa-ode} for TPA, but for the time-averaged fluence instead of the instantaneous intensity.
Consequently, the photoconductivity profile in a medium with an ESA or SA nonlinearity will also be described by Eq.~\eqref{eq:sigma-tpa-profile}, but with different expressions for $\ftpa$.

\section{General formalism for the optical coefficients}
\label{sec:general-formalism}
To derive the optical reflection and transmission coefficients for the photoconductivity profiles in Sec.~\ref{sec:photoconductivity-models}, we turn to the theory of electromagnetic waves in stratified media~\cite{born2019, aspnes1973, wait1970, ginzburg1964, lekner2016}.
Although this is an old and well-developed topic in optics, we review it here for completeness, to establish common notation, and to comment on aspects of the problem that are important for the solutions that we give in Sec.~\ref{sec:analytic-solutions}.

\subsection{The Maxwell equations in a stratified medium}
\label{sec:maxwell}
\begin{figure}[tbp]
\begin{center}
\includegraphics[width=\columnwidth]{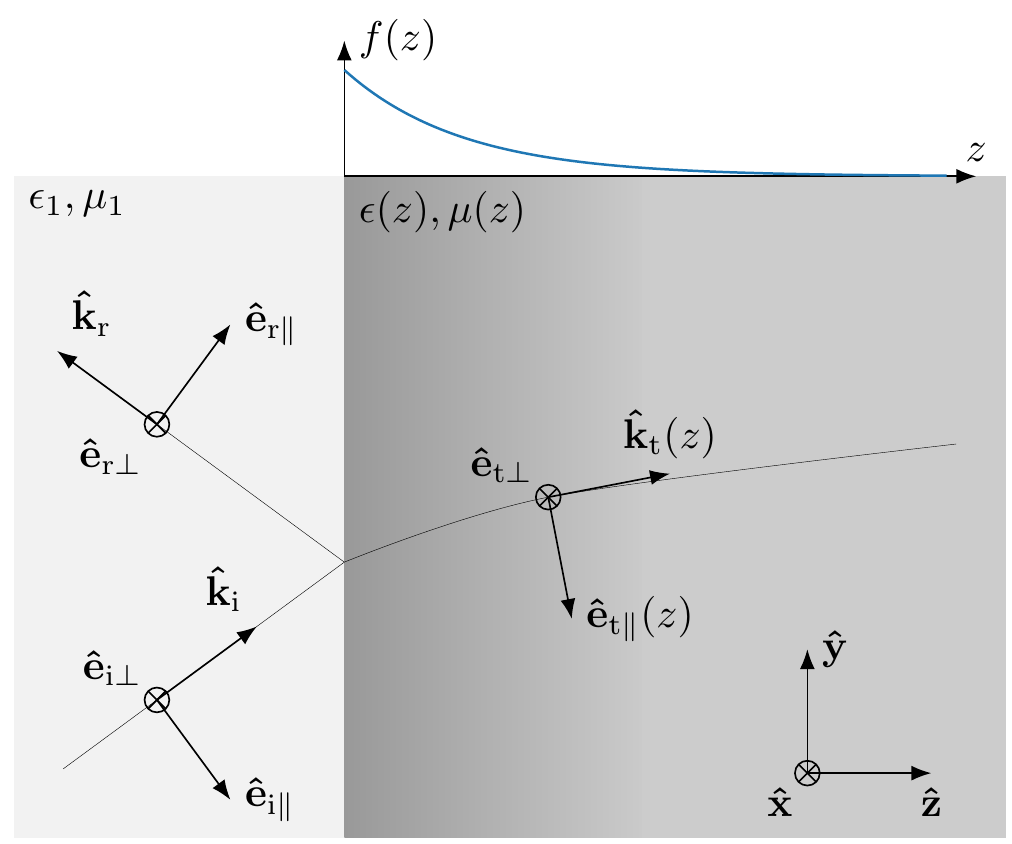}
\caption{Coordinate systems for reflection and transmission of an electromagnetic wave incident on a stratified medium.}
\label{fig:coordinates}
\end{center}
\end{figure}

Following Born and Wolf~\cite{born2019}, we consider a monochromatic wave with angular frequency $\omega$ traveling in the $yz$-plane through a stratified medium with relative permittivity \mbox{$\epsr(z) = \epsilon(z)/\epsilon_0$} and relative permeability \mbox{$\mur(z) = \mu(z)/\mu_0$}, as shown in Fig.~\ref{fig:coordinates}.
The s-polarized (TE) and p-polarized (TM) fields may be expressed in terms of the dimensionless functions $\Us$, $\Vs$, $\Up$ and $\Vp$,
\begin{align}
\begin{gathered}
\begin{aligned}
E_x &= \mathfrak{E}\Us(z)e^{i \beta k_0 y},%
&Z_0 H_y &= \mathfrak{E}\Vs(z)e^{i \beta k_0 y},%
\end{aligned}\\%
Z_0 H_z = -\beta \mathfrak{E}\frac{\Us(z)}{\mur(z)}e^{i \beta k_0 y}; \label{eq:scaled-fields-s}%
\end{gathered}\\[1ex]%
\begin{gathered}%
\begin{aligned}%
Z_0 H_x &= \mathfrak{E}\Up(z)e^{i \beta k_0 y},%
&E_y &= -\mathfrak{E}\Vp(z)e^{i \beta k_0 y},%
\end{aligned}\\%
E_z = \beta\mathfrak{E}\frac{\Up(z)}{\epsr(z)}e^{i \beta k_0 y}.
\label{eq:scaled-fields-p}%
\end{gathered}%
\end{align}
where \mbox{$k_0 = \omega/c$}, $Z_0$ is the impedance of free space, $\mathfrak{E}$ is a characteristic field scale, and $\beta$ is a constant that determines the propagation direction.
For a wave traveling in a medium with a local refractive index \mbox{$n = \sqrt{\epsr\mur}$} at an angle $\theta$ with respect to the $z$-axis, \mbox{$\beta = n\sin\theta$}.

Substituting Eq.~\eqref{eq:scaled-fields-s} and Eq.~\eqref{eq:scaled-fields-p} into the Maxwell curl equations, we obtain two systems of first-order equations governing electromagnetic wave propagation in the medium,
\begin{align}
\dv{z}\begin{bmatrix} \Us \\ \Vs \end{bmatrix} &= i k_0\begin{bmatrix} 0 & \mur \\
\epsr - \beta^2/\mur & 0 \end{bmatrix}\begin{bmatrix} \Us \\ \Vs \end{bmatrix};\label{eq:ode-system-s}\\[1ex]
\dv{z}\begin{bmatrix} \Up \\ \Vp \end{bmatrix} &= i k_0\begin{bmatrix} 0 & \epsr \\ \mur - \beta^2/\epsr & 0 \end{bmatrix}\begin{bmatrix} \Up \\ \Vp \end{bmatrix}.\label{eq:ode-system-p}
\end{align}
In the limit as \mbox{$z \to \infty$}, where the perturbation vanishes, these ODEs yield the harmonic solutions 
\begin{align}
\begin{bmatrix} \Us \\ \Vs \end{bmatrix} &= A_+\begin{bmatrix} 1 \\ \eta/\mur \end{bmatrix}e^{i\eta k_0 z} + A_-\begin{bmatrix} 1 \\ -\eta/\mur \end{bmatrix}e^{-i\eta k_0 z};\\
\begin{bmatrix} \Up \\ \Vp \end{bmatrix} &= A_+\begin{bmatrix} 1 \\ \eta/\epsr \end{bmatrix}e^{i\eta k_0 z} + A_-\begin{bmatrix} 1 \\ -\eta/\epsr \end{bmatrix}e^{-i\eta k_0 z},
\end{align}
where \mbox{$\eta^2 = \epsr\mur - \beta^2$} and $A_+$ and $A_-$ are the complex amplitudes for waves travelling in the forward and backward directions, respectively.
The first-order ODEs in Eqs~\eqref{eq:ode-system-s} and \eqref{eq:ode-system-p} can be written as second-order ODEs by eliminating $\Vs$ and $\Vp$,
\begin{align}
& \dv[2]{\Us}{z} + k_0^2(\mur \epsr - \beta^{2})\Us = 0;%
\label{eq:waveeqgen-s}\\
& \dv[2]{\Up}{z} + k_0^2(\mur \epsr - \beta^{2})\Up = 0.%
\label{eq:waveeqgen-p}
\end{align}
 
\subsection{Reflection and transmission coefficients}
\label{sec:reflection-transmission}
Now consider a stratified half-space, \mbox{$z\ge 0$}, with light incident at an angle $\theta_1$ from a homogeneous medium with permittivity $\epsilon_1$, permeability $\mu_1$, refractive index \mbox{$n_1 = c\sqrt{\epsilon_1\mu_1}$}, and characteristic impedance \mbox{$Z_1 = \sqrt{\mu_1/\epsilon_1}$}.
Phase continuity requires that \mbox{$\beta = n_1\sin\theta_1$} remains constant across the boundary, so if we define $\theta_2$ to be the angle that the transmitted wave makes with the $z$-axis at $z=0$, \mbox{$\beta = \sqrt{\epsr(0)\mur(0)}\sin\theta_2$}.
It is also useful to define the complex field ratios
\begin{align}
\Phis(z) &= \frac{\Vs}{\Us} = \frac{1}{ik_0\mur}\frac{\Us'}{\Us} = \frac{1}{ik_0\mur}\frac{d\ln\Us}{dz}; \label{eq:s-field-ratio}\\
\Phip(z) &= \frac{\Vp}{\Up} = \frac{1}{ik_0\epsr}\frac{\Up'}{\Up} = \frac{1}{ik_0\epsr}\frac{d\ln\Up}{dz}.
\label{eq:p-field-ratio}
\end{align}

Applying the boundary condition that the tangential components of the electric field and magnetic field remain continuous across the interface, we can express the reflection and transmission coefficients at the boundary, \mbox{$z=0$}, as ~\cite{born2019}
\begin{align}
\rs &= \frac{Z_0\cos\theta_1 - Z_1\Phis(0)}{Z_0\cos\theta_1 + Z_1\Phis(0)},\\
\ts &= \frac{2Z_0\cos\theta_1}{Z_0\cos\theta_1 + Z_1\Phis(0)}; \label{eq:r-t-general-s}\\
\rp &= \frac{Z_1\cos\theta_1 - Z_0\Phip(0)}{Z_1\cos\theta_1 + Z_0\Phip(0)},
\intertext{and}
\tp &= \frac{\cos\theta_1}{\cos\theta_2}\,\frac{2Z_0\Phip(0)}{Z_1\cos\theta_1 + Z_0\Phip(0)} \label{eq:r-t-general-p}.
\end{align}
Using Eqs.~\eqref{eq:r-t-general-s} and \eqref{eq:r-t-general-p}, we can calculate the exact expressions for the reflection amplitude and the transmission for a given $\epsr(z)$ using the forward-travelling solutions to the wave equation evaluated at the interface, \mbox{$z = 0$}.

Following \textcite{vakhnenko1970}, it is useful to express the reflection and transmission amplitudes in terms of the perturbation parameter $\Sigma$,
\begin{align}
r &= \frac{\bar{r} - \Sigma}{1 + \Sigma}; \label{eq:r_Sigma} \\
t &= \frac{\bar{t}}{1 + \Sigma}, \label{eq:t_Sigma}
\end{align}
where $\bar{r}$ and $\bar{t}$ are the reflection and transmission amplitudes of an unperturbed homogeneous medium with permittivity $\epsilon_2$~\cite{vakhnenko1970}.
Expressing the perturbed coefficients this way has the important advantage that \mbox{$\lim_{\Sigma\rightarrow\infty}r = 1$}.
We have also found that it provides smaller approximation errors than a more conventional perturbative series~\cite{lekner2016}.

We can now write $\Sigma$ for both polarizations in terms of the complex field ratios in Eq.~\eqref{eq:s-field-ratio} and Eq.~\eqref{eq:p-field-ratio}:
\begin{align}
\Sigma_{\text{s}} &= \frac{Z_{1}}{Z_{0}}\frac{\bar{Z}_{2} \Phis(0) - Z_{0}\mathrm{cos}\bar{\theta}_{2}}{\bar{Z}_{2}\mathrm{cos}\theta_{1} + Z_{1}\mathrm{cos}\bar{\theta}_{2}}; \label{eq:Sigma_spol}
 \\ \Sigma_{\text{p}} &= \frac{\bar{Z}_{2} \cos\theta_{2} - Z_{0} \Phip(0)}{Z_{1} \cos\,\theta_{1} + \bar{Z}_{2}\cos\bar{\theta}_{2}},
 \label{eq:ppol_Sigma}
\end{align}
where $\bar{Z}_{2}$ and $\bar{\theta}_{2}$ are the equilibrium impedance and transmission angle, respectively.
These expressions enable us to compute the perturbation parameter if we know the fields at the surface. In Sec.~\ref{sec:solutions-maxwell}, below, we discuss methods for determining these fields.

\subsection{Solutions to the Maxwell equations}
\label{sec:solutions-maxwell}
For an arbitrary $\epsr(z)$, one typically uses numerical methods to solve Eqs.~\eqref{eq:ode-system-s} and \eqref{eq:ode-system-p} to determine the field ratios $\Phis$ and $\Phip$, respectively, at the interface.
A common approach is to use the transfer matrix method, where the inhomogeneously perturbed medium is treated as a series of homogeneous layers with constant permittivity~\cite{abeles1950, born2019}.
Alternatively, one can use the Runge-Kutta or other methods to find the solution with a standard ODE solver.
In both approaches, an initial condition for the fields must be set at a location $z = z_i$ where the perturbation is small, then back-propagated to the interface at $z = 0$. The fields at $z_i$ correspond to the forward-traveling harmonic solution in the unperturbed medium.
However, these approaches are prone to both numerical and systematic errors, which can arise because the differential equation is stiff or because the initial fields can not be specified with sufficient precision. These problems are especially important when the equilibrium medium is absorbing, since any error in the initial condition will blow up exponentially as they approach the interface. 

Alternatively, in Sec.~\ref{sec:analytic-solutions} we will show that Eq.~\eqref{eq:waveeqgen-s} has analytic solutions for the main four depth profiles discussed in Sec.~\ref{sec:photoconductivity-models}.
With these solutions, we may derive analytic expressions for the optical coefficients by directly evaluating the field ratio $\Phis(0)$.
Compared to numerical solutions, analytic expressions make it easier to identify the key parameters that control the coefficients.
We may then readily determine controlled approximations for the optical coefficients that are valid in different limits of these parameters.
Finally, these analytic solutions may also be used to validate numerical solutions in parameter ranges that are numerically challenging.

\section{Analytic solutions}
\label{sec:analytic-solutions}
Here we derive analytic solutions to the wave equation for the photoconductivity depth profiles introduced in Sec.~\ref{sec:photoconductivity-models}, derive expressions for the optical coefficients, and discuss various limiting forms of these coefficients.
We treat the exponential model in Sec.~\ref{sec:exponential-profile}, the saturable profile and the TPA profile in Sec.~\ref{sec:epstein-profile-solutions}, and the combined TPA-saturable profile in Sec.~\ref{sec:combined-profile-solutions}.

\subsection{Exponential Profile}
\label{sec:exponential-profile}
As we noted in Sec.~\ref{sec:Introduction}, the exponential profile is the default assumption for analyzing pump--probe spectroscopy measurements.
Multiple authors have discussed the solutions to the wave equation for this profile and applied it in different contexts~\cite{vakhnenko1970, gallant1982, vinet1984, sabbah2002, meng2015}.
Although our primary interest is in nonlinear photoconductivity profiles, a thorough understanding of this linear profile is valuable for identifying signatures of nonlinearity, and we are unaware of a systematic exploration of the solution in all of its physical limits.
Consequently, we review the solution here and discuss a variety of approximations to it that are valid in different physical limits.
This discussion will set the stage for a similar analysis of the nonlinear profiles in Secs.~\ref{sec:epstein-profile-solutions} and \ref{sec:combined-profile-solutions}.

When expressed in terms of the relative permittivity, the exponential profile has the form
\begin{equation}
\epsr(z) = \bar{n}^2 + \varrho^2 e^{-\alpha z},
\end{equation}
where $\bar{n}$ is the equilibrium refractive index and \mbox{$\varrho^2 = i\chif F/(\omega\epsilon_0)$} is the photoinduced permittivity change at the surface.
Letting \mbox{$\eta^2 = \bar{n}^2 - \beta^2$}, the wave equation for an $s$-polarized probe is
\begin{equation}
\dv[2]{\Us}{z} + k_0^2(\eta^2 + \varrho^2 e^{-\alpha z})\Us = 0.
\label{eq:waveeqexp}
\end{equation}

Substituting \mbox{$k_0^2\eta^2 = -\alpha^2\nu^2/4$}, \mbox{$k_0^2\varrho^2e^{-\alpha z} = \alpha^2\xi^2/4$}, we can transform Eq.~\eqref{eq:waveeqexp} into the Bessel equation~\cite[Eq.~14.1.3.2]{polyanin2017},
\begin{gather}
\xi^2\dv[2]{\tilde{\Us}}{\xi} + \xi\dv{\tilde{\Us}}{\xi} + (\xi^2 - \nu^2)\tilde{\Us} = 0,
\label{eq:waveeqbessel}
\end{gather}
where \mbox{$\tilde{U_\text{s}}[(2 k_0\varrho/\alpha)\mathrm{exp} (-\alpha z/2)] = U_\text{s}(z)$}.
The two linearly independent solutions to the above equation are~\cite{wait1952}
\begin{equation}
\tilde{\Us}(\xi) = C_1 J_{-\nu}(\xi) + C_2 J_\nu(\xi),
\end{equation}
which can be expressed in terms of the variables in Eq.~\eqref{eq:waveeqexp} as
\begin{align}
\Us(z) &= C_1 J_{-2i\eta k_0/\alpha}\left[(2k_0\varrho/\alpha)\mathrm{exp} (-\alpha z/2) \right] \nonumber\\
&\qquad + C_2 J_{2i\eta k_0/\alpha}\left[(2k_0\varrho/\alpha)\mathrm{exp} (-\alpha z/2)\right].\label{eq:wavesolexp}
\end{align}
We can associate these solutions with waves traveling forward and backward through the medium.
In the limit that the perturbation has fully decayed, \mbox{$\alpha z\rightarrow\infty$}, we can use \mbox{$\lim_{\xi\rightarrow 0}J_{\nu}\left(\xi\right)\sim(\xi/2)^{\nu}/\Gamma\left(\nu+1\right)$} to recover the conventional harmonic solutions to the wave equation for the equilibrium response~\cite[Eq.~10.7.3]{olver2024},
\begin{align}
\lim_{\alpha z\rightarrow\infty}\Us &= K_1 e^{i\eta k_0 z} + K_2 e^{-i\eta k_0 z}.
\end{align}
Note that the positive root in the harmonic solution is associated with the negative order of the Bessel function in Eq.~\eqref{eq:wavesolexp}, and vice versa.

\subsubsection{Reflection and transmission amplitude}
\label{sec:reflection-exponential}
As we discussed in Sec.~\ref{sec:reflection-transmission}, we may write both $r$ and $t$ for a perturbed medium in terms of the perturbation parameter $\Sigma$, given in Eq.~\eqref{eq:Sigma_spol}, which in turn depends on complex field ratio, \mbox{$\Phi_\text{s}(z = 0)$}, defined in Eq.~\eqref{eq:s-field-ratio}.
For the forward-traveling solution in Eq.~\eqref{eq:wavesolexp}, we have
\begin{equation}
\Phi_\text{s}(0) = i \varrho \frac{J'_{-\nu}(\xi_0)}{J_{-\nu}(\xi_0)},
\label{eq:phi_exp}
\end{equation}
where \mbox{$\xi_0 = 2k_0\varrho/\alpha$}.
Applying the recurrence relation~\cite[Eq.~10.6.2]{olver2024}
\begin{equation}
J'_{\nu}(\xi)= J_{\nu-1}(\xi) - (\nu/z)J_{\nu}(\xi)
\end{equation} 
and substituting the result into Eq.~\eqref{eq:Sigma_spol}, we obtain
\begin{equation}
\Sigma = -i \frac{J_{- \nu + 1}(\xi_0)}{J_{-\nu}(\xi_0)} \frac{\varrho}{n_1 \cos\theta_{1} + \bar{n} \cos\bar{\theta}_{2}}.
\label{eq:Sigma_exp}
\end{equation}

We can associate the quantities \mbox{$\nu = 2ik_0\eta\Lambda$} and \mbox{$\xi_0 = 2k_0\varrho\Lambda$} with dimensionless ratios of the relevant length scales of the problem.
First, we can write \mbox{$\nu \propto \Lambda/\delta_{\eta}$}, where $\Lambda$ is the pump attenuation length and \mbox{$\delta_{\eta} = \sqrt{2/{\mu_0 \bar{\sigma} \omega}}$} has the form of the skin depth for the probe in the equilibrium medium, with conductivity $\bar{\sigma}$.
Second, we can write $\xi_0\propto \Lambda/\delta_{\varrho}$, where \mbox{$\delta_{\varrho} = \sqrt{2/{\mu_0 \dsigmasurf \omega}}$} has the form of the skin depth for the probe in a fictitious medium with conductivity $\dsigmasurf$.
We note here that as defined, both $\delta_{\eta}$ and $\delta_{\varrho}$ are complex, so they are not skin depths, in the strict sense.
Nonetheless, the association provides useful intuition about the physical behavior in different regimes of $\nu$ and $\varrho$, so we adopt it with this caveat.

\subsubsection{Typical parameter values}
\label{sec:typical-parameter-values}
In Table~\ref{tab:parameters}, we present the magnitude of $\nu$ and $\xi_0$ for several pump-probe experiments, along with $\bar{\sigma}$ and $\dsigmasurf$ at the probe frequency, the wavelengths of the pump and probe fields, the static permittivity $\epsilon_\text{s}$, and the pump attenuation length $\Lambda$.
We use \mbox{$n_{1} = 1$} and \mbox{$\theta_{1} = 0$} to standardize the calculation for $\nu$ and $\xi_0$.
Several of the examples in Table~\ref{tab:parameters} are from reports of photoinduced superconductivity that show evidence of saturation, but as we will see in Sec.~\ref{sec:epstein-profile-solutions}, these parameters are relevant for the saturation profile, also.
Since our purpose is to discuss qualitative trends in $\nu$ and $\xi_0$, we do not account for the systematic error in $\dsigmasurf$ that saturation may induce in these experiments.

\begin{table*}[t]
  \centering
  \caption{Experimental parameters for various pump-probe measurements}
  \label{tab:parameters}
  \begin{ruledtabular}
  \begin{tabular}{ccccccccccc}
    Material & Pump $\lambda$ &  Probe $\lambda$ & $\bar{\sigma}$  & $\dsigmasurf$ & $\epsilon_\text{s}$ & $\Lambda$ & $\delta_{\eta}$ & $\delta_{\varrho}.$ & $|\nu|$ & $|\xi_0|$ \\
    & ($\mu$m) & ($\mu$m) & $(\Omega^{-1}\,\text{cm}^{-1})$ & $(\Omega^{-1}\,\mathrm{cm}^{-1})$ & & ($\mu$m) & ($\mu$m) & ($\mu$m) & & \\
    YBa$_{2}$Cu$_{3}$O$_{6.5}$\footnotemark[1] & 0.8  & 294 & $15 - 10i$ & $4 + 15i$ & 4.5 & 0.2 & 19 & 9 &  0.05 & 0.05 \\
    
    LBCO\footnotemark[2] & 0.8  & 300 & $2.2 - 13i$ & $1 + 6i$ & 4.5 & 0.4 & 120 & 15 & 0.6 & 0.05 \\

    YBa$_{2}$Cu$_{3}$O$_{6.5}$\footnotemark[1] & 1.42  & 294 & $15 - 10i$ & $1.1 + 7.2i$ & 4.5 & 0.6 & 19 & 13 & 0.15 & 0.09 \\
   
    Au\footnotemark[3] & 2.95 & 3.65 & $2790 + 31400i$ & $137 - 114i$ & 7.24 & 0.02 & 0.02 & 0.7 & 1.15 & 0.09 \\
    
    YBa$_{2}$Cu$_{3}$O$_{6.5}$ \footnotemark[1]& 16  & 294 & $15 - 10i$ & $4 + 8i$ & 4.5 & 0.7 & 19 & 12 & 0.2 & 0.1 \\
    
    LBCO\footnotemark[2] & 2 & 300 & $2.2 - 13i$ & $0.06 + 1.4i$ & 4.5 & 2.5 & 130 & 30 & 0.6 & 0.2 \\
    GaAs\footnotemark[4] & 0.79 & 298 & --- & $21 + 22i$ & --- & 0.41 & 85 & 7 & 0.06 & 0.13 \\
    
    K$_{3}$C$_{60}$\footnotemark[5] & 7.3 & 190 & $240 + 350i$ & $-230 + 250i$ & 5 & 0.22 & 1.45 & 1.65 & 0.31 & 0.28 \\
    
    YBa$_{2}$Cu$_{3}$O$_{6.5}$\footnotemark[1] & 10  & 294 & $15 - 10i$ & $0.4 + 1i$ & 4.5 & 2.4 & 19 & 32 & 0.5 & 0.15\\
    
    YBa$_{2}$Cu$_{3}$O$_{6}$\footnotemark[6] & 0.5  & 0.73 & $892 - 91i$ & $-137 + 343i$ & 4.5 & 0.01 & 0.14 & 0.01 & 2.43 & 1.23 \\
    (insulating) &  &  & & & & & & & & \\
    \text{BEDT-TTF}\footnotemark[7] & 8 & 200 & $168 - 18i$ & $-108 + 215i$ & 5 & 3 & 3.4 & 2 & 2.65 & 3.15 \\
  \end{tabular}
    \end{ruledtabular}
    \footnotetext[1]{\textcite{liu2020a}}
    \footnotetext[2]{\textcite{casandruc2015}}
    \footnotetext[3]{\textcite{Sielcken2020}}
    \footnotetext[4]{\textcite{beard2000}}
    \footnotetext[5]{\textcite{budden2021}}
    \footnotetext[6]{\textcite{sahota2019}}
    \footnotetext[7]{\textcite{buzzi2020}}
\end{table*}

\subsubsection{Limiting forms}

While Eq.~\eqref{eq:Sigma_exp} is valid for all $\nu$ and $\xi_0$, Table~\ref{tab:parameters} shows that their magnitudes are typically between 0.01 and 10.
Here, we derive approximations for $r$ and $t$ that span these limits.

\emph{Weak perturbation limit.}
We associate this limit with \mbox{$\dsigmasurf \rightarrow 0$}, which implies \mbox{$\delta_{\varrho} = \sqrt{2/(\mu_0 \dsigmasurf \omega)} \rightarrow \infty$}.
For constant $k_0 \Lambda$, we also have \mbox{$\xi_0 \propto \Lambda/\delta_{\varrho} \rightarrow 0$}.
To calculate the limiting form of $\Sigma$ in this case, consider the Taylor expansion of the Bessel function around \mbox{$\xi = 0$},
\begin{equation}
J_{\nu}(\xi) = \left( \frac{\xi}{2} \right)^{\nu} \sum_{k = 0}^{\infty} (-1)^{k} \frac{(\xi/4)^{2k}}{k! \Gamma(\nu + k + 1)}.
\label{eq:taylor-bessel}
\end{equation}
Retaining only the zeroth-order term in the series and using the recurrence relation for the $\Gamma$-function,
\begin{equation}
\Gamma (\nu + 1) = \nu \Gamma (\nu),
\label{eq:gamma_recurrence}
\end{equation}
we can write the ratio of the Bessel functions in Eq.~\eqref{eq:Sigma_exp} as
\begin{equation}
\frac{J_{1 - \nu}(\xi_{0})}{J_{- \nu}(\xi_{0})} \approx \frac{\xi_{0}/2}{1- \nu}.
\label{eq:bessel-ratio-1}
\end{equation}
Substituting Eq.~\eqref{eq:bessel-ratio-1} in Eq.~\eqref{eq:Sigma_exp}, we obtain
\begin{equation}
\begin{split}
\lim_{\dsigmasurf \rightarrow 0}\Sigma &= \frac{-i k_0 \varrho^{2}\Lambda}{\left(1 - 2ik_0\bar{n}\Lambda \right) (n_1 + \bar{n})} \\
&= \frac{\dsigmasurf \mathrm{Z}_0 \Lambda}{\left(1 - 2ik_0\bar{n}\Lambda \right) (n_1 + \bar{n})},
\end{split}
\label{eq:Sigma-weak-perturbation}
\end{equation}
where we assume \mbox{$\theta_{1} = \theta_{2} = 0$} and use \mbox{$\varrho^{2} = i \dsigmasurf/ \omega \epsilon_0$}.
Substituting Eq.~\eqref{eq:Sigma-weak-perturbation} in Eq.~\eqref{eq:r_Sigma} and Eq.~\eqref{eq:t_Sigma}, we obtain
\begin{equation}
\lim_{\dsigmasurf \rightarrow 0}r = \frac{n_1 - \bar{n} - \dsigmasurf Z_{0}\Lambda/(1 - 2ik_0\bar{n}\Lambda)}{n_1 + \bar{n} + \dsigmasurf Z_{0}\Lambda/(1 - 2ik_0\bar{n}\Lambda)},
\label{eq:r-weak-perturbation}
\end{equation}
and 
\begin{equation}
\lim_{\dsigmasurf \rightarrow 0}t = \frac{2 n_1}{n_1 + \bar{n} + \dsigmasurf Z_{0}\Lambda/(1 - 2ik_0\bar{n}\Lambda)}.
\label{eq:t-weak-perturbation}
\end{equation}

\emph{Long-wavelength limit.}
We associate this limit with \mbox{$k_0\Lambda \rightarrow 0$}, which implies both \mbox{$\delta_{\eta}\rightarrow\infty$} and \mbox{$\delta_{\varrho}\rightarrow\infty$} because \mbox{$\delta_{\eta},\delta_{\varrho} \propto 1/\sqrt{\omega} \propto 1/\sqrt{k_0}$}.
Consequently, \mbox{$\xi_0 \propto \Lambda/\delta_{\varrho} \rightarrow 0$} and \mbox{$\nu \propto \Lambda/\delta_{\eta} \rightarrow 0$}.
In these limits, we can neglect the term \mbox{$\nu = 2ik_0\bar{n}\Lambda \propto \Lambda/\delta_{\eta}$} in Eq~\eqref{eq:Sigma-weak-perturbation} to obtain
\begin{align}
\lim_{k_0\Lambda \rightarrow 0}\Sigma &= \frac{-ik_0\varrho^2\Lambda}{n_1 + \bar{n}}
= \frac{\dsigmasurf  Z_{0}\Lambda}{n_1 + \bar{n}}.
\label{eq:Sigma-tinkham}
\end{align}
The reflection and transmission coefficients are then
\begin{equation}
\lim_{k_0\Lambda \rightarrow 0}r = \frac{n_{1} - \bar{n} - \dsigmasurf  Z_{0}\Lambda}{n_1 + \bar{n} + \dsigmasurf  Z_{0}\Lambda}
\label{eq:r_tinkham}
\end{equation}
and 
\begin{equation}
\lim_{k_0\Lambda \rightarrow 0}t = \frac{2n_1}{n_1 + \bar{n} + \dsigmasurf  Z_{0}\Lambda}.
\label{eq:t_tinkham}
\end{equation}
Equation~\eqref{eq:t_tinkham} often goes by the name of the Tinkham formula, which Glover and Tinkham derived for far-infrared transmission spectroscopy~\cite{glover1957}.

By retaining the first two terms of the series in Eq.~\eqref{eq:taylor-bessel} and following a similar procedure to the one that led to Eq.~\eqref{eq:bessel-ratio-1}, we can derive the first correction to the long-wavelength limit.
This gives
\begin{equation}
\frac{J_{- \nu + 1}(\xi_0)}{J_{- \nu}(\xi_0)} = \frac{\xi_0/2}{- \nu+1} \frac{\left[1 - \frac{\xi_0^{2}/4}{- \nu + 2} \right]}{\left[1 - \frac{\xi_0^{2}/4}{- \nu + 1} \right]} + \mathcal{O}(\xi_0^3).
\end{equation}
Substituting this expression in Eq.~\eqref{eq:Sigma_exp} and setting \mbox{$\Lambda/\delta_{\eta} \rightarrow 0$}, we obtain
\begin{align}
\Sigma &=  \frac{\dsigmasurf Z_{0} \Lambda}{n_1 + \bar{n}} \left[ 1 + \frac{i}{2}\mu_0\dsigmasurf\omega\Lambda^2 \right] + \ldots,
\label{eq:Sigma_tinkham-first-o}
\end{align}
which includes a correction of order $(\Lambda/\delta_\varrho)^2$ to the phase shift.

\emph{High-photoconductivity limit.}
We associate this limit with \mbox{$\dsigmasurf \rightarrow \infty$}, which implies \mbox{$\delta_{\varrho} \rightarrow 0$} and, in turn, \mbox{$\Lambda/\delta_{\varrho} \rightarrow \infty$}.
We can deriive the limiting form of Eq.~\eqref{eq:Sigma_exp} for this case cby considering the asymptotic limit of the Bessel function,
\begin{gather}
\lim_{\xi\rightarrow\infty}J_{\nu}(\xi) = \sqrt{\frac{2}{\pi \xi}}\,\mathrm{cos}\left(\xi - \frac{1}{2} \nu \pi - \frac{1}{4} \pi\right).
\label{eq:bessel-ratio}\\
\intertext{In this limit,}
lim_{\xi_0 \rightarrow \infty} \frac{J_{- \nu + 1}(\xi_0)}{J_{- \nu}(\xi_0)} = \lim_{\xi_0 \rightarrow \infty} \mathrm{tan}\left(\xi_0 + \frac{1}{2} \nu \pi - \frac{1}{4} \pi \right)\approx i.
\label{eq:bessel_ratio_asym}
\end{gather}
Substituting Eq.~\eqref{eq:bessel_ratio_asym} in Eq.~\eqref{eq:Sigma_exp}, we obtain
\begin{equation}
\lim_{\xi_0 \rightarrow \infty}\Sigma = \frac{\varrho}{n_1 + \bar{n}} = \frac{\sqrt{i \dsigmasurf/ \omega \epsilon_0}}{n_1 + \bar{n}}.
\label{eq:Sigma-asym}
\end{equation}
The expressions for $r$ and $t$ are
\begin{equation}
\lim_{\xi_0 \rightarrow \infty}r = \frac{n_1 - \bar{n} - \sqrt{i \dsigmasurf/ \omega \epsilon_0}}{n_1 + \bar{n} + \sqrt{i \dsigmasurf/ \omega \epsilon_0}},
\label{eq:r_asym}
\end{equation}
and
\begin{equation}
\lim_{\xi_0 \rightarrow \infty}t = \frac{2n_1}{n_1 + \bar{n} + \sqrt{i \dsigmasurf/ \omega \epsilon_0}}.
\label{eq:t_asym}
\end{equation}

\emph{Short-wavelength or bulk limit.}
We associate this limit with \mbox{$k_0\Lambda \rightarrow \infty$}, which implies the divergence of both the order of the Bessel function, \mbox{$\nu \propto \Lambda/\delta_{\eta}$}, and its argument, \mbox{$\xi_0 \propto \Lambda/\delta_{\varrho}$}.
To simplify Eq.~\eqref{eq:Sigma_exp} in this limit, it is convenient to use the following identity~\cite[Eq.~10.10.1]{olver2024},
\begin{align}
\lim_{k_0\Lambda \rightarrow \infty}f &= \frac{J_{-\nu +1}(\xi)}{J_{-\nu}(\xi)}\\
&= \cfrac{1}{-2\nu \xi^{-1}-\cfrac{1}{-2\nu \xi^{-1}-\cfrac{1}{-2\nu \xi^{-1}-\cdots}}}\\
&= \frac{1}{-2\nu \xi^{-1} - f},
\end{align}
which yields a quadratic for $f$ with roots
\begin{equation}
f = -\nu \xi^{-1} \pm \sqrt{\nu^2 \xi^{-2} - 1}.
\end{equation}
The negative root of this expression is unphysical, as it corresponds to choosing the negative root of the photoexcited permittivity to obtain its refractive index.
Choosing the positive root and substituting into Eq.~\eqref{eq:Sigma_exp}, we obtain
\begin{equation}
\begin{split}
\lim_{k_0\Lambda \rightarrow \infty}\Sigma &= \left(\sqrt{\bar{n}^2 + i\frac{\dsigmasurf}{\omega\epsilon_0}} - \bar{n}\right)\frac{1}{n_1 + \bar{n}}\\
&= \frac{n_\text{ex} - \bar{n}}{n_1 + \bar{n}},
\end{split}
\label{eq:Sigma-bulk}
\end{equation}
where \mbox{$n_\text{ex} = \sqrt{\bar{n}^2 + i\dsigmasurf/(\omega\epsilon_0)}$} is the refractive index of the photoexcited medium at the surface.
Substituting Eq.~\eqref{eq:Sigma-bulk} into Eqs.~\eqref{eq:r_Sigma} and~\eqref{eq:t_Sigma}, and simplifying, we obtain
\begin{align}
\lim_{k_0\Lambda \rightarrow \infty}r &= \frac{n_1 - n_\text{ex}}{n_1 + n_\text{ex}}
\label{eq:r_exp_bulk}
\intertext{and}
\lim_{k_0\Lambda \rightarrow \infty}t &= \frac{2 n_1}{n_1 + n_\text{ex}},
\label{eq:t_exp_bulk}
\end{align}
which are simply the Fresnel reflection and transmission amplitudes for a homogeneously medium with refractive index $n_\text{ex}$.
\begin{figure}[tbp]
\begin{center}
\includegraphics[width = \columnwidth]{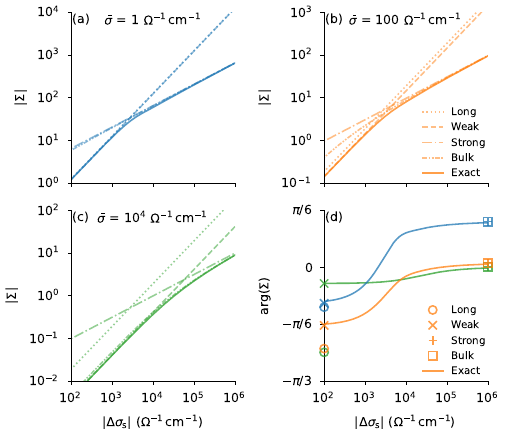}
\caption{%
Magnitude (a--c) and phase (d) of $\Sigma$ for an exponential profile as a function of $|\dsigmasurf|$, for three values of $\sigmabar$ that are displayed in (a--c) and indicated by the associated color or grayscale in (d).
Throughout, $k_0= 0.01~\text{m}^{-1}$ and $\Lambda = 1~\mu\text{m}$.
The exact result for $\Sigma$ in Eq.~\eqref{eq:Sigma_exp} is shown together with four different limiting approximations, indicated in the legend of (b) and discussed in the text.
Markers in (d) show only the limiting values of the phase given by the four approximations, as indicated in that legend.
}
\label{fig:Exp-sigma}
\end{center}
\end{figure}

Figure~\eqref{fig:Exp-sigma} shows the magnitude and phase of $\Sigma$ as a function of $|\dsigmasurf|$, along with the limiting forms described above.
\mbox{Figure~\eqref{fig:Exp-sigma}(a)--(c)} show the magnitude for different values of $\bar{\sigma}$, while Figure~\eqref{fig:Exp-sigma}(d) shows the phase for the same values of $\bar{\sigma}$ as in \mbox{Figure~\eqref{fig:Exp-sigma}(a)--(c)}.
When $|\dsigmasurf|$ is small, the weak-perturbation limit provides a good approximation to both the magnitude and phase of $\Sigma$, with $|\Sigma|\propto|\dsigmasurf|$..
When both $|\dsigmasurf|$ and $\bar{\sigma}$ are small, the long wavelength limit provides a good approximation for $|\Sigma|$, but $\arg(\Sigma)$ shows some discrepancy.
As the skin depth of the probe field decreases with increasing $\bar{\sigma}$ in panels (a)--(c), the parameter \mbox{$\nu \propto \Lambda/\delta_{\eta}$} increases for constant $\Lambda$, causing $|\Sigma|$ to deviate from the long wavelength limit.

The behavior of $|\Sigma|$ changes from \mbox{$|\Sigma|\propto|\dsigmasurf|$} to \mbox{$|\Sigma|\propto\sqrt{|\dsigmasurf|}$} as $|\dsigmasurf|$ increases, with the transition region moving towards higher values of $|\dsigmasurf|$ as $\sigmabar$ increases.
Physically, this crossover occurs when the skin depth associated with the photoexcitation, \mbox{$\delta_{\varrho}\propto1/\sqrt{\dsigmasurf}$}, becomes comparable to $\Lambda$ or $\delta_\nu\propto1/\sqrt{\sigmabar}$, whichever is smaller.
In Fig.~\ref{fig:Exp-sigma}(a), where $\sigmabar$ is smallest, $\delta_\nu \gg \Lambda$ and the transition with $\dsigmasurf$ occurs when $\delta_{\varrho}\approx\Lambda$.
In Fig.~\ref{fig:Exp-sigma}(c), where $\sigmabar$ is largest, $\delta_\nu \lesssim \Lambda$, so the relevant comparison becomes $\delta_{\varrho}\approx\delta_{\nu}$.
In this case, the probe pulse is effectively sampling a uniformly excited region, since the probe field is localized to the photoexcitation depth $\Lambda$, and the bulk-limit approximation is good for all values of $\dsigmasurf$.
Figure~\ref{fig:Exp-sigma}(d) shows that the phase of $\Sigma$ is relatively constant on either side of the transition region, and that it matches the limiting approximations in all cases except the long-wavelength limit, which requires that both $\delta_\nu\ll\Lambda$ and $\delta_\varrho\ll\Lambda$.
Although the phase error is relatively small for $\sigmabar = 1~\Omega^{-1}\,\text{cm}^{-1}$, where $\delta_\nu \gg \Lambda$, it increases with $\sigmabar$ to exceed 0.5 when $\sigmabar = 10^4~\Omega^{-1}\,\text{cm}^{-1}$, where $\delta_\nu \lesssim \Lambda$.
This discrepancy highlights the superiority of the weak-perturbation approximation over the more commonly used long-wavelength approximation.

\subsection{Nonlinear profiles}
\label{sec:epstein-profile-solutions}
Having thoroughly studied the properties of the optical coefficients for the exponential photoconductivity profile, we are now in a position to examine the nonlinear profiles that we introduced in Sec.~\ref{sec:photoconductivity-models}.
We will start with the saturable profile from Sec.~\ref{sec:saturable-photoconductivity} and the TPA profile from Sec.~\ref{sec:tpa-photoconductivity}, then turn to the combined saturation-TPA profile from Sec.~\ref{sec:combined-photoconductivity}.
We show that the saturation profile and the TPA profile are special cases of a more general profile, introduced by Epstein nearly 100 years ago, which yields wave solutions that are described by hypergeometric functions~\cite{epstein1930, ginzburg1964, lekner2016}. We also show that the combined profile yields solutions described by Heun functions, which appear in several other areas of physics~\cite{ronveaux1995, olver2024}.

\subsubsection{General solutions for the saturable and TPA profiles}
\label{sec:gen-sol-sat-tpa}
With some changes in notation to adapt it to our context, we write the Epstein profile as
\begin{equation}
\epsilon_{r} = \bar{n}^{2} + \varsigma^{2}\frac{e^{- \alpha (z - z_0)}}{1 + e^{- \alpha (z - z_0)}} + \vartheta^{2}\frac{e^{- \alpha (z - z_0)}}{[1 + e^{- \alpha (z - z_0)}]^{2}}.
\label{eq:epsilon_epstein}
\end{equation}
We obtain the saturable profile, Eq.~\eqref{eq:sigma-sat-profile}, with the substitutions
\begin{gather}
\begin{gathered}
\begin{aligned}
\vartheta &= 0, &\varsigma^{2} &= i\dsigmasat/(\omega\epsilon_0) & e^{\alpha z_0} &= F/\fsat,
\end{aligned}\\
\end{gathered}
\label{eq:epstein-sat-substitutions}
\intertext{and we obtain the TPA profile, Eq.~\eqref{eq:sigma-tpa-profile}, with the substitutions}
\begin{gathered}
\begin{aligned}
\vartheta^{2} &= -i\dsigmatpa/(\omega\epsilon_0), &\varsigma^{2} &= 0,
\end{aligned}\\
e^{\alpha z_0} = -(F/\ftpa)/(1 + F/\ftpa).
\end{gathered}
\label{eq:epstein-tpa-substitutions}
\ \end{gather}
We may now solve the wave equation for both Eq.~\ref{eq:sigma-sat-profile} and Eq.~\eqref{eq:sigma-tpa-profile} by solving it for Eq.~\eqref{eq:epsilon_epstein}~\cite{epstein1930, ginzburg1964, lekner2016}, then making the substitutions in Eqs.~\eqref{eq:epstein-sat-substitutions} and \eqref{eq:epstein-tpa-substitutions}, respectively.

To proceed with the solution, we substitute Eq.~\eqref{eq:epsilon_epstein} into the wave equation, Eq.~\eqref{eq:waveeqgen-s}, with the additional substitutions \mbox{$u = e^{-\alpha (z - z_0)} = u_0 e^{-\alpha z}$}, \mbox{$\tilde{U}_\text{s}(u) = U_\text{s}(z)$}, and \mbox{$\kappa = k_0\Lambda$}, to put it in the form 
\begin{align}
&u^{2}\frac{d^{2}\tilde{U}_\text{s}}{du^{2}} + u\frac{d\tilde{U}_\text{s}}{du} \notag \\ & + \kappa^{2}\left[\eta^{2} + \varsigma^{2}\frac{u}{1 + u} + \vartheta^{2}\frac{u}{(1 + u)^{2}}\right]\tilde{U}_\text{s} = 0.
\end{align}
This has the form of Riemann's differential equation with thee regular singularities at \mbox{$u = 0, -1, \infty $}~\cite[Eq.~15.11.1]{olver2024}.
For the singularity at $u = 0$, the two independent solutions for $U_\text{s}$ are
\begin{widetext}
\begin{align}
U_{\text{s}1} (z) &=  u^{a} (1 + u)^{b} F (a + b + c, a + b - c; 1 + 2a; -u ) \\
U_{\text{s}2} (z) &=  u^{-a} (1 + u)^{b} F (-a + b + c, -a + b - c; 1 - 2a; -u ),
\label{eq:solepstein}
\end{align}
\end{widetext}
where
\begin{equation}
\begin{gathered}
\begin{aligned}
a^{2} &= - \kappa^{2} \eta^{2}, &b(b-1) &= - \kappa^{2}\vartheta^{2},
\end{aligned}\\
c^{2} = - \kappa^{2} (\eta^{2} + \varsigma^{2}),
\end{gathered}
\end{equation}
and $F$ is the hypergeometric function. To identify the forward-travelling solution, we evaluate the limit $z\rightarrow\infty$, which corresponds to $u\rightarrow 0$.
Choosing the root $a = i\kappa\eta$, the asymptotic solutions are then
\begin{align}
\lim_{z\rightarrow\infty}U_{\text{s}1} (z) &= u^a = e^{-ik_0\eta(z-z_0)},\\
\lim_{z\rightarrow\infty}U_{\text{s}2} (z) &= u^{-a} = e^{ik_0\eta(z-z_0)},
\end{align}
and we see that Eq.~\eqref{eq:solepstein} for $U_{\text{s}2}$ is the forward-travelling solution.

To determine the perturbation parameter $\Sigma$, we substitute Eq.~\eqref{eq:solepstein} in Eq.~\eqref{eq:s-field-ratio},
\begin{equation}
\Phi(0) = -\frac{i}{\kappa} \left\lbrace a - \frac{b u_0}{u_0 + 1} + u_0 \frac{(a - b)^{2} - c^{2}}{1 - 2a} \psi(0) \right\rbrace,
\end{equation}
where
%\begin{widetext}
\begin{equation}
\squeezespaces{0.25}
\psi(0) = \frac{F(-a + b + c + 1, -a - b -c + 1; -2a + 2; -u_0)}{F(-a + b + c, -a + b -c ; -2a + 1; -u_0)}.
\end{equation}
%\end{widetext}
This enables us to write
\begin{equation}
\label{eq:epstein-Sigma}
\squeezespaces{0.25}
\Sigma = \frac{i}{k_0\Lambda (n_{i} \cos \theta_{i} + \eta)} \left\lbrace \frac{b u_0}{u_0 + 1} - u_0 \frac{(a - b)^{2} - c^{2}}{1 - 2a} \psi(0) \right\rbrace.
\end{equation}
To obtain the optical coefficients for the saturable and TPA profiles, we can substitute Eqs.~\eqref{eq:epstein-sat-substitutions} and \eqref{eq:epstein-tpa-substitutions}, respectively, into Eq.~\eqref{eq:epstein-Sigma}.
We then substitute Eq.~\eqref{eq:epstein-Sigma} into Eqs.~\eqref{eq:r_Sigma} and \eqref{eq:t_Sigma} to obtain $r$ and $t$, respectively. This completes the general solution for the optical coefficients for these two profiles.

\subsubsection{Long-wavelength approximations for the saturable and TPA profiles}
\label{sec:long-wave-sat-tpa}
The expression for $\Sigma$ in Eq.~\eqref{eq:epstein-Sigma} is considerably more complex than the equivalent expression in Eq.~\eqref{eq:Sigma_exp} for the exponential profile, and we have not explored its limits at the same level of detail.
Nonetheless, we can derive expressions for the long-wavelength limit, \mbox{$\kappa = k_0\Lambda \rightarrow 0$}, which are equivalent to ones we derived with more physical arguments in Ref.~\cite{dodge2023}.
As we indicated in Table~\ref{tab:parameters}, experiments are commonly in this limit, so exact results that describe it are valuable.

To proceed, we note that for \mbox{$\kappa \rightarrow 0$}, the parameters of Eq.~\eqref{eq:epstein-Sigma} become \mbox{$a = i \kappa \eta$}, \mbox{$b = \kappa^{2} \vartheta^{2}$}, and \mbox{$c = i \kappa \sqrt{\eta^{2} + \varsigma^{2}}$}. We also have~\cite[Eq.~15.2.1]{olver2024}
\begin{multline}
\lim_{\kappa\rightarrow 0}F(-a+b+c, -a+b-c; 1-2a; -u_0) \\
= 1 - \frac{(a-b)^2-c^2}{1-2a}u_0 = 1
\end{multline}
and~\cite[Eq.~15.4.1]{olver2024}
\begin{multline}
\lim_{\kappa\rightarrow 0}F(-a+b+c+1, -a+b-c+1; 2-2a; -u_0) \\
= F(1, 1; 2; -u_0) = u_0^{-1}\ln(1 + u_0).
\end{multline}
Substituting in Eq.~\eqref{eq:epstein-Sigma}, we get
\begin{equation}
\squeezespaces{0.25}
    \lim_{\kappa\to0} \Sigma =  \frac{-i\kappa}{(n_{i} \cos \theta_{i} + \eta)} \left[ \varsigma^2 \ln (1 + e^{-\alpha z_0}) + \frac{\vartheta^2 e^{\alpha z_0}}{1+ e^{\alpha z_0}} \right],
\end{equation}
which we can use to determine $\Sigma$ for both the saturable profile and the TPA profile.

Substituting Eq.~\eqref{eq:epstein-sat-substitutions} for the saturable profile, this becomes
\begin{align}
\lim_{k_0\Lambda\to0} \Sigma &= \frac{\dsigmasat Z_0\Lambda \ln (1+F/\fsat)}{n_1 \mathrm{cos}\theta_{1} + \eta}.
\label{eq:sat-chi_thin}
\end{align}
At normal incidence, this yields
\begin{equation}
\lim_{\kappa\to0} r = \frac{n_1 - \bar{n} - \dsigmasat \Lambda Z_0 \ln (1 + F/\fsat)}{n_1 + \bar{n} + \dsigmasat \Lambda Z_0 \ln (1 + F/\fsat)}
\label{eq:sat-r_thin}
\end{equation}
and
\begin{equation}
\lim_{\kappa\to0} t = \frac{2 n_1}{n_1 + \bar{n} +  \dsigmasat \Lambda Z_0 \ln (1 + F/\fsat)}.
\label{eq:sat-t_tinkham}
\end{equation}
Alternativelly, substituting Eq.~\eqref{eq:epstein-tpa-substitutions} for the TPA profile yields
\begin{equation}
\lim_{k_0\Lambda\to0} \Sigma = \frac{\dsigmatpa \Lambda Z_0 F/\ftpa}{n_{1}\cos \theta_{1} + \eta},
\label{eq:tpa_chi-thin}
\end{equation}
which for normal incidence gives
\begin{align}
\lim_{\kappa\to0} r &= \frac{1 - \bar{n} - \dsigmatpa \Lambda Z_0 F/\ftpa}{1 + \bar{n} + \dsigmatpa \Lambda Z_0 F/\ftpa}
\label{eq:tpa-r-tinkham}
\intertext{and}
\lim_{\kappa\to0} t &= \frac{2 n_1}{n_1 + \bar{n} +  \dsigmatpa \Lambda Z_0F/\ftpa}.
\label{eq:tpa-t_tinkham}
\end{align}
As we discussed in Ref.~\cite{dodge2023}, the optical coefficients for the saturable profile have a logarithmic dependence on fluence.
This is because as the conductivity saturates near the surface, the depth of the saturated region grows logarithmically, increasing its effective thickness.
By contrast, the optical coefficients for the TPA profile are proportional to the fluence in the long wavelength limit.
In this case, although the surface photoconductivity has a superlinear dependence on fluence, there is a compensating effect on the effective thickness of the profile~\cite{dodge2023}.
Consequently, the optical response with a TPA nonlinearity is indistinguishable from that of the exponential profile.

\subsection{Combined Profile}
\label{sec:combined-profile-solutions}
In this section, we derive the solutions to the wave equation in the presence of both TPA/ESA and saturation.
The wave equation is
\begin{equation}
    \frac{d^{2} U_\text{s}}{dz^{2}} + k_0^{2} \left[\eta^{2} + i\frac{\dsigma(z, F; \fsat, \ftpa)}{\omega\epsilon_0} \right]U_\text{s} = 0,
\label{eq:comb-wave-eq}
\end{equation}
where \mbox{$\dsigma(z, F; \fsat, \ftpa)$} is given in Eq.~\eqref{eq:sigma-tpa-sat-profile}.
Substituting \mbox{$u = e^{- \alpha z}$} and \mbox{$\tilde{U}_\text{s}(u) = U_\text{s}(z)$}, we can put Eq.~\eqref{eq:comb-wave-eq} in the form
\begin{equation}
\squeezespaces{0.5}
    \frac{d^{2} \tilde{U}_\text{s}}{du^{2}} + \frac{1}{u}\frac{d \tilde{U}_\text{s}}{du} + 
    \left[ \frac{\kappa^{2}\eta^{2}}{u^{2}} + \frac{Q(F)}{u (u - u_{+}) (u - u_{-})} \right]\tilde{U}_\text{s} = 0,
\label{eq:comb-diff-eq}
\end{equation}
where $\kappa = k_0\Lambda$,
\begin{equation}
    Q(F) = \kappa^{2}\chi_{F} F_{\text{TPA}} \frac{\left(F_{\text{TPA}} - F\right)}{F}
\end{equation}
and $u_{\pm}$ correspond to the solutions of
\begin{multline}
    u^{2} + \left[ \left( F_{\text{TPA}}/F_{\text{sat}} - 2 \right) \left( 1 +  F_{\text{TPA}}/F_{\text{sat}}\right) \right] u \\
    + \left( F_{\text{TPA}}/F_{\text{sat}} + 1\right)^{2} = 0,
\end{multline}\\
\begin{multline}
    u_{\pm} = ( 1 + F_{\text{TPA}}/F) \Bigl[ ( 1 - F_{\text{TPA}}/2F_{\text{sat}} ) \\
    \pm  ( F_{\text{TPA}}/2F_{\text{sat}}) \sqrt{1 - 4F_{\text{TPA}}/F_{\text{sat}}}\Bigr].
\end{multline}
Equation~\eqref{eq:comb-diff-eq} has four regular singularities at $0, u_{\pm}$, and $\infty$, which enables us to solve it by transforming it into the Heun differential equation~\cite{olver2024, ronveaux1995},
\begin{multline}
\frac{d^2w}{d\zeta^2} + \left(\frac{\gamma}{\zeta} + \frac{\delta}{\zeta-1} + \frac{\epsilon}{\zeta - a}\right)\frac{dw}{d\zeta} \\
+ \frac{\alpha\beta\zeta - q}{\zeta(\zeta - 1)(\zeta - a)}w = 0,
\label{eq:heun}
\end{multline}
with
\begin{equation}
\gamma + \delta + \epsilon = \alpha + \beta + 1
\label{eq:heun-constraint}
\end{equation}
and singularities at 0, 1, $a$, and $\infty$

To proceed with the transformation, we solve the indicial equation at each of the singularities to obtain $\rho = \pm i\kappa\eta$ for the singularities at 0 and $\infty$, and $\rho = \{0, 1\}$ for the singularities at $u_\pm$. We can then move the singularity at $u_+$ to 1 by changing variables to \mbox{$\zeta = u/u_{+}$}, \mbox{$\hat{U}_\text{s}(\zeta) = \tilde{U}_\text{s}(u)$}
\begin{equation}
    \frac{d^{2} \hat{U}_\text{s}}{d\zeta^{2}} + \frac{1}{\zeta}\frac{d \hat{U}_\text{s}}{d\zeta} + 
    \left[ \frac{\kappa^{2}\eta^{2}}{\zeta^{2}} + \frac{Q(F)/u_+}{\zeta (\zeta - 1) (\zeta - a)} \right]\hat{U}_\text{s} = 0,
\label{eq:comb-diff-eq-zeta}
\end{equation}
which also moves the singularity at $u_{-}$ to
\begin{equation}
a = \frac{u_{-}}{u_{+}} = \left(1 - \frac{\ftpa}{2\fsat} - \frac{\ftpa}{2\fsat}\sqrt{1 - \frac{4\fsat}{\ftpa}}\right)^2.
\end{equation}

Lastly, we write $\hat{U}_\text{s} = \zeta^{-i\kappa\eta}w(\zeta)$, which transforms Eq.~\eqref{eq:comb-diff-eq-zeta} into the standard form given in Eq.~\eqref{eq:heun},
\begin{equation}
    \frac{d^{2} w}{d\zeta^{2}} + \frac{1-2i\kappa\eta}{\zeta}\frac{d w}{d\zeta} + 
    \frac{Q(F)/u_{+}}{\zeta (\zeta - 1) (\zeta - a)} w = 0.
\label{eq:comb-diff-eq-w}
\end{equation}
with $\gamma = 1 - 2i\kappa\eta$, $\delta=0$, $\epsilon=0$, and \mbox{$q = Q(F)/u_{+}$}. To specify $\alpha$ and $\beta$, we solve the incidial equation for the singularity at 0 to get $\rho = \{0, 2i\kappa\eta\}$ and for the singularity at $\infty$ to get $\rho = \{0, -2i\kappa\eta\}$. The constraint in Eq.~\eqref{eq:heun-constraint} then requires that $\alpha=0$ and $\beta = -2i\kappa\eta$.

With the parameters of the standard form established, we may write the solution to Eq.~\eqref{eq:comb-diff-eq} in terms of the local Heun function~\cite{olver2024, ronveaux1995},
\begin{equation}
\hat{U}_\text{s} = \zeta^{- i \kappa \eta} Hl (a, q; 0, -2i\kappa \eta, 1 - 2i\kappa \eta, 0, \zeta).
\label{eq:combined-solution}
\end{equation}
We can confirm that this is the forward-traveling solution by evaluating the limit $z\to \infty$, equivalent to $\zeta\to 0$~\cite[Eq.~31.3.1]{olver2024},
\begin{equation}
\lim_{\zeta\rightarrow 0}\hat{U}_\text{s} = \zeta^{- i \kappa \eta} = e^{ik_0\eta z}.
\end{equation}
In principle, we may now obtain the optical coefficients for the combined saturation-TPA profile in the same way that we have for the exponential, saturable, and TPA profiles.
In practice, this is much more challenging in this case than it is with the others.
Substituting Eq.~\eqref{eq:combined-solution} into Eq.~\eqref{eq:s-field-ratio} does not readily yield a form that allow further progress, so we will defer further development in this direction until a later investigation.
We do note, however, that we are unaware of any other example of a stratified medium profile with wave solutions described by Heun functions, so the one presented here may be of interest to others who study the properties of these functions.

\section{Conclusion}
\label{sec:conclusion}
We have described a general framework for determining the optical coefficients of a photoexcited medium, and we applied it to specific examples of photoconductivity depth profiles that are likely to occur in experiments.
We discussed several different profiles, including ones that show a nonlinear dependence on incident fluence.
In addition to profiles that have been discussed previously, we developed a new, two-parameter family that describes how a medium will respond when saturation and TPA nonlinearities are present simultaneously.
We showed that the wave equation for s-polarized probe waves has analytic solutions for many of the profiles that we discussed, one of which appears to have not been discussed previously.
We derived approximations to these solutions and analyzed their limits of validity in terms of the natural length scales involved in pump-probe experiments.
In subsequent work, we will discuss how one may use the results presented here to improve the uncertainty in pump-probe measurements of the surface conductivity.

\begin{acknowledgments}
J.\ S.\ D.\ acknowledges support from NSERC and CIFAR, and D.\ G.\ S.\ from an NSERC Alexander Graham Bell Canada Graduate Scholarship.
\end{acknowledgments}

\end{document}